# DYNAMICAL MODELLING OF
# HOT STELLAR SYSTEMS

David Merritt

*Department of Physics and Astronomy, Rutgers University,*
*Piscataway, NJ 08855 USA*

**Abstract.** Estimation of the distribution function $f$ and potential $\Phi$ of hot stellar systems from kinematical data is discussed. When the functional forms of $f$ and $\Phi$ are not specified *a priori*, accurate estimation of either function requires very high quality data : either accurate "line profiles" at radii extending well beyond an effective radius, or large samples ($N \gtrsim 10^3$) of discrete radial velocities. Estimates of $\Phi(r)$ based on much smaller data sets can be very strongly influenced by assumptions, explicit or implicit, about the form of $f$. The importance of casting the estimation problem into a mathematically determined form is stressed. Some techniques for nonparametric estimation are presented, with some preliminary results of their application to real stellar systems.

## 1. INVERSE PROBLEMS

The problem of inferring the dynamical state of a hot stellar system like an elliptical galaxy from kinematical observations is an example of what statisticians call "inverse problems." The simplest sort of inverse problem consists of finding a function $f$ that appears inside of an integral :

$$g(x) = \int_a^b k(x,y)f(y) \; dy, \qquad (1)$$

with $g(x)$ and $k(x,y)$ known functions. For instance, $f$ might represent the intrinsic luminosity distribution of a star on the sky, $k$ the smearing effect of the earth's atmosphere, and $g$ the observed image. However statisticians define inverse problems more broadly, to include any problem that requires "making inferences about a phenomenon from partial or incomplete information" (O'Sullivan 1986). This definition includes statistical estimation – that is, estimating the parameters of a function given a sample drawn from that function – as well as model building. They further divide the class of inverse problems into two types, "well-posed" and "ill-posed." Well-posed problems have – at least mathematically – unique solutions, and those solutions are possible to find, in an approximate sense, even when the information is incomplete or imperfect. Much more common are ill-posed inverse problems : problems that have no unique solution, even given perfect or near-perfect data. The most common ill-posed problems are underdetermined ones; examples are finding the axis ratio distribution of triaxial galaxies from the distribution of Hubble types, calculating the distribution function of a spherical galaxy from its density profile, and inferring the 3-D velocity field of a triaxial galaxy from line-of-sight velocities. To the extent that we see galaxies in projection, almost every attempt at model building begins as an underdetermined problem.

However, even problems that are mathematically fully determined can be ill-posed, in the sense that the solution is extremely sensitive to errors or incompleteness in the data. The instability can arise in a number of ways. For instance, in the integral equation (1), we can

imagine adding to the true solution $f(y)$ a term of the form $A\cos(my)$; since

$$\int_a^b k(x,y)\cos(my)dy \to 0 \qquad \text{as} \quad m \to \infty \qquad (2)$$

for any integrable kernel $k(x,y)$ (the so-called "Riemann-Lebesgue theorem"), the added term will contribute negligibly to $g(x)$, regardless of the amplitude of $A$, for large enough $m$. This means that high-frequency components in the data, whether real or due to noise, become amplified in the solution, and the problem becomes more severe as the solution grid is made finer (Phillips 1962). Even when the data are perfect, additional errors resembling noise are liable to be introduced in the course of computation. the the to largassociated with stfunction $g(x)$ corresp$f(y)$. Crudely speaking, features in the "source" function $f(y)$ whose extent is comparable to or smaller than the smoothing scale are impossible to recover, and even features that are more extended will often be masked by spurious oscillations in the numerical solution.

A second class of ill-posed problems arise when one wishes to estimate the form of an unknown frequency function $f(x)$ given a discrete sample $x_j$, $j = 1, ..., n$, drawn from that function. Here a standard approach is to represent the unknown function $f$ through a set of parameters $\theta_i$, and to vary the parameters until the "likelihood" $L$ is maximized, where

$$L = \prod_{j=1}^n f(x_j|\theta_i). \qquad (3)$$

However if we imagine making the number of parameters $\theta_i$ very large – or equivalently, if we place no prior constraints on the form of $f$ (aside from very general conditions such as integrability and positivity) – it is obvious that the "most likely" solution will always look something like

$$f(x) = \frac{1}{n}\sum_j \delta(x - x_j), \qquad (4)$$

a set of delta functions at the data points. Thus maximum likelihood estimation, like the inversion of an ill-conditioned integral equation, tends inevitably to amplify fluctuations in the data (Tapia & Thompson 1978). This problem is really a ubiquitous one when estimating the form of a function nonparametrically : in the absence of prior information, probability theory will always lead to a frequency function for which every minute detail of the data was inevitable. Although maximum likelihood estimates are often described as having the quality of being "smooth," this is only true when the number of parameters $\theta_i$ is small compared to the number of data points, i.e. when the form of $f$ is determined more by preconceptions of the scientist than by the data.

Dynamical modelling of hot stellar systems always involves elements of both deconvolution and estimation. For instance, the intrinsic density profile of a star cluster is related to its surface density through Abel's integral, and that surface density must be inferred from the positions of individual stars. Because these problems of deprojection and estimation are intrinsically unstable, one would expect galactic astronomers to have accumulated a great deal of experience with ill-posed problems, and with the techniques that have been developed for dealing with them. Certainly this is true in most other fields where inverse problems are prominent, such as geophysics, meteorology, crystallography, etc. By and large, however, astronomers have tended to view inverse problems as if they had well-defined solutions. For instance, the standard technique for deprojecting a surface brightness profile is to fit the data to a smooth function (e.g. a Hubble or $r^{1/4}$ law) with only two or three parameters, and invert this function analytically. Over time, these fitting functions tend to acquire the status of physical laws, in spite of inevitable cases where they don't fit terribly well.

There are a number of problems with modelling galaxies in this way. Fitting a smooth function to the data is tantamount to "adjusting" the data points by small amounts so that they lie along a smooth curve. But if the inverse problem is ill-conditioned, small adjustments in the data can translate into large, and unpredictable, variations in the solution. This means that two galaxies whose surface brightness profiles are well fit by the same function might have rather different intrinsic density profiles. A more fundamental objection to the use of *ad hoc* functions is that there is a natural tendency to interpret the fitting parameters as physically meaningful quantities, even in cases where the available data do not contain enough information to uniquely determine those quantities. For instance, the dynamical properties of globular clusters (central mass density, velocity anisotropy, etc.) are usually equated to the corresponding parameters in the best-fitting Michie-King model, even though surface brightness and velocity dispersion profiles typically impose only order-of-magnitude constraints on such quantities, regardless of the quality of the data (Merritt 1992b).

Even when the inverse problem is mathematically fully determined, the range of possible solutions given by a parametric fit will always be smaller than that given by a nonparametric algorithm; thus, parametrically derived confidence intervals will always be too small, sometimes drastically so. Finally, while simple parametrized functions may be adequate for the description of functions of a single variable, such as surface brightness, it is much harder to guess the appropriate form of functions that describe data sets of two or more dimensions.

A standard technique for dealing nonparametrically with ill-conditioned inverse problems is the "method of regularization" (Phillips 1962; Turchin et al. 1971; Miller 1974; Tikhonov & Arsenin 1977; Craig & Brown 1986). Instead of attempting an exact inversion, one chooses a solution that minimizes a weighted combination of two functionals. The first functional measures the deviation of the model from the data and the second measures the "implausibility" of the model. For instance, if $\nu_p(r_p)$ is the observed surface density of a galaxy at projected radius $r_p$, and $\nu(r)$ is the (unknown) intrinsic density profile, one could choose to minimize

$$\sum_{data} \left[ \frac{\nu_p - \mathbf{A}\nu}{\Delta \nu_p} \right]^2 + \lambda P(\nu). \tag{5}$$

Here $\mathbf{A}$ represents the projection operator, and $\Delta \nu_p$ is the estimated uncertainty in the measured $\nu_p$. The functional $P$ is chosen so that "implausible" solutions have large values, and the scalar $\lambda$ determines the degree of smoothing. Similarly, when estimating the density profile from a discrete set of positions, one could maximize the "penalized likelihood," whose logarithm is

$$\sum_{data} \log \left[ \mathbf{A}\nu \right] (r_p) - \lambda P(\nu), \tag{6}$$

subject to the constraint that the integrated number $\int \nu(r)d^3r$ equal the number of stars in the sample. For obscure reasons, $P(\nu)$ is called a "regularization functional" in the context of ill-posed integral equations (e.g. Craig & Brown 1986), and a "penalty function" in the context of estimation problems (e.g. Tapia & Thompson 1978).

Needless to say, there is considerable art involved in the selection of $P$ and $\lambda$. Two general philosophies are widely adhered to in the choice of $P$. When the data are too sparse to specify a unique solution, or when the inverse problem is mathematically underdetermined, one chooses $P$ so that – in the limit of infinite $\lambda$, i.e. infinite "smoothing" – the solution assumes some reasonable form. For instance, setting

$$P(f) = \int_{-\infty}^{\infty} \left[ (d/dx)^3 \log f(x) \right]^2 dx \qquad (7)$$

produces an "infinitely smooth" solution that is a Gaussian, with mean and standard deviation determined by the data (Silverman 1982). Similarly, choosing $P(f) = f - f^*$ yields a smooth solution close to $f^*$. When the inverse problem is mathematically determined, but ill-conditioned, one is worried primarily about spurious high-frequency oscillations; thus $P$ should be chosen to be large whenever the solution is rapidly varying. A common choice is

$$P(f) = \int w(x) \left( d^n f/dx^n \right)^2 dx. \qquad (8)$$

Here $w(x)$ a weighting function, which for strongly inhomogeneous solutions should scale something like $1/f^2$, while $n$ is the "order" of the regularization; the larger $n$, the more sensitive $P$ will be to small fluctuations in $f$.

Whatever the choice of smoothing functional $P$, an increase in $\lambda$ produces an increase in deviations of the model from the data and a decrease in spurious fluctuations. One generally chooses $\lambda$ to be as large as possible without forcing the solution to differ significantly (in a $\chi^2$ sense, say) from the data. Practice suggests that, when the order of the regularization is sufficiently large, even small amounts of smoothing can produce spectacular reductions in the high-frequency components of the solution (Titterington 1985). Thus – at least in the case of mathematically *determined* inverse problems – the choice of smoothing functional $P$ is often not crucial : any functional that is sensitive to rapid variations in $f$ will do. For underdetermined problems, however, the solution will generally depend strongly on $P$.

The method of regularization consists essentially in replacing an ill-posed inverse problem by a stable minimization problem. Alternative schemes for dealing with ill-posed problems have been discussed, including expansion of the solution in terms of a truncated set of basis functions (e.g. Fricke 1952; Baker 1977; Dejonghe 1989a), and iterating from an initial, smooth guess (Landweber 1951; Richardson 1972; Lucy 1974). In general, these alternative methods are judged inferior to regularization because of their limited flexibility (Miller 1974). For instance, in a basis function expansion, the degree of smoothness depends in a complicated way on the choice of basis set and the number of terms retained.

## 2. ABEL INVERSIONS

The solution of Abel's integral equation provides a good example of an ill-posed inverse problem. Consider a spherical galaxy with intrinsic density profile $\nu(r)$ and projected surface density

$\nu_p(r_p)$. We have

$$\nu_p(r_p) = \int_{r_p^2}^{\infty} \frac{\nu(r)dr^2}{\sqrt{r^2 - r_p^2}} \tag{9}$$

with exact solution

$$\nu(r) = -\frac{1}{\pi} \int_r^{\infty} \frac{d\nu_p}{dr_p} \frac{dr_p}{\sqrt{r_p^2 - r^2}}. \tag{10}$$

In texts on integral equations, the operation on the right hand side of equation (10) is called a "differentiation of order 1/2" of $\nu_p$ with respect to $r_p$ (e.g. Sneddon 1972, p. 272). This means that the Abel equation is rather mildly ill conditioned, compared to equations whose solution involves the equivalent of full or even multiple derivatives of the data. Nevertheless, a formal solution like equation (10) is of limited use when dealing with real data.

A standard technique for the numerical solution of integral equations like equation (9) is "product integration." First replace the integration by a discrete sum :

$$\nu_p(r_p) = \int_{r_p^2}^{\infty} \frac{\nu(r)dr^2}{\sqrt{r^2 - r_p^2}} \approx \sum_{j=1}^{n} \nu(r_j) \int_{r_{j-1}^2}^{r_j^2} (r^2 - r_p^2)^{-1/2} dr^2$$

$$\approx 2 \sum_{j=1}^{n} \nu(r_j) \left[ \sqrt{r_j^2 - r_p^2} - \sqrt{r_{j-1}^2 - r_p^2} \right], \tag{11}$$

then write

$$\nu_p(r_{p_i}) = \sum_{j=1}^{n} A_{ij} \nu(r_j),$$

$$A_{ij} = 2 \left[ \sqrt{r_j^2 - r_{p_i}^2} - \sqrt{r_{j-1}^2 - r_{p_i}^2} \right], \tag{12}$$

a matrix equation which can be solved by standard techniques. The result (e.g. Gorenflo & Kovetz 1966) is generally a solution in which fluctuations in the data are amplified to an unacceptable level. Following the discussion above, we might try to increase the smoothness by minimizing the quantity

$$\sum_{i=1}^{m} \left[ \frac{\nu_p(r_{p_i}) - \sum_{j=1}^{n} A_{ij} \nu_j}{\Delta \nu_p} \right]^2 + \lambda \int_0^{\infty} \left[ \frac{1}{\nu_0} \frac{d^2 \nu}{dr^2} \right]^2 dr$$

$$\approx \sum_{i=1}^{m} \left[ \frac{\nu_p(r_{p_i}) - \sum_{j=1}^{n} A_{ij} \nu_j}{\Delta \nu_p} \right]^2 + \lambda \sum_{j=1}^{n} \left( \frac{1}{\nu_{0j}} \right)^2 \frac{[\nu_{j+1} - 2\nu_j + \nu_{j-1}]^2}{(r_j - r_{j-1})^3}. \tag{13}$$

(Note that a uniform grid in $r$ has been assumed.) This is "second order regularization," with weighting function $\nu_0(r)$. By choosing a specified function $\nu_0(r)$ to provide the weighting – rather than $\nu(r)$ itself – the quantity to be minimized is a quadratic expression in the $\nu_j$, which means that a standard least-squares or quadratic programming algorithm can be used. (The latter choice permits the easy imposition of other constraints, e.g. positivity or monotonicity of $\nu(r)$; see Dejonghe 1989a.) Figure 1 shows the regularized deprojection of simulated surface brightness measurements drawn, with 10% errors, from the projected Plummer law; $\nu_0(r)$ was taken to be the intrinsic Plummer density profile. The deprojected density profile of Figure

1d is quite close to the "correct" one, except for a slight central cusp – a reminder that the behavior of the density very near the center is impossible to determine uniquely from limited data, even when the "seeing" is perfect.

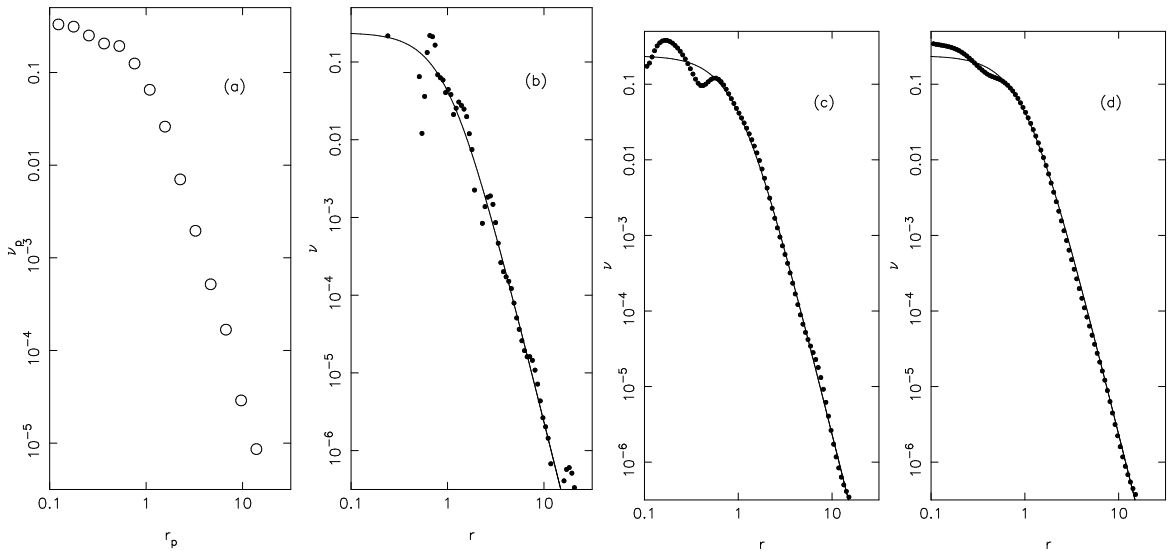

Fig. 1. (a) Random realization of a Plummer surface density law, with 20% "errors." (b)-(d) : Regularized deprojections; $\lambda$ increases by about four orders of magnitude between frames. In (b), many of the points are at negative values of $\nu$.

A similar algorithm can be used to estimate the form of the density profile nonparametrically from a set of discrete, projected positions. Here the quantity to be maximized is

$$\prod_{i=1}^{m} \nu_p(r_{p_i}) \exp\left[-\lambda P(\nu)\right], \qquad (14)$$

the "penalized likelihood" (Tapia & Thompson 1978, p. 102), or equivalently

$$\sum_{i=1}^{m} \log \sum_{j=1}^{n} A_{ij}\nu_j - \lambda \sum_{j=1}^{n} \left(\frac{1}{\nu_j}\right)^2 \frac{\left[\nu_{j+1} - 2\nu_j + \nu_{j-1}\right]^2}{\left(r_j - r_{j-1}\right)^3}. \qquad (15)$$

Because the likelihood is already a strongly nonlinear function of $\nu(r)$, there is no computational disadvantage in setting the weighting function in $P(\nu)$ to $1/\nu^2$, rather than $1/\nu_o^2$ as in the example above. This choice leads to a "smoothness" criterion that is completely independent of any prior notions about the form of $\nu(r)$. Minimization can be achieved with a general nonlinear optimization algorithm such as BCONF of IMSL or E04JAF of NAG. The minimization must be carried out subject to the constraint that the total number of stars implied by $\nu(r)$ equal the number in the sample; this constraint may be imposed via an additional penalty function, of the form

$$\delta \left[4\pi \int_0^\infty \nu(r)r^2\, dr - N_{data}\right]^2, \qquad (16)$$

or by using a general "nonlinear programming" algorithm (e.g. NCONF of IMSL) and fixing the total number via an additional linear constraint. Figure 2 shows the density profile estimated from a sample of 300 stars drawn from the same Plummer model in Figure 1. The regularized estimator works very well, yielding a smooth profile that closely approximates the true one, even for this modest sample.

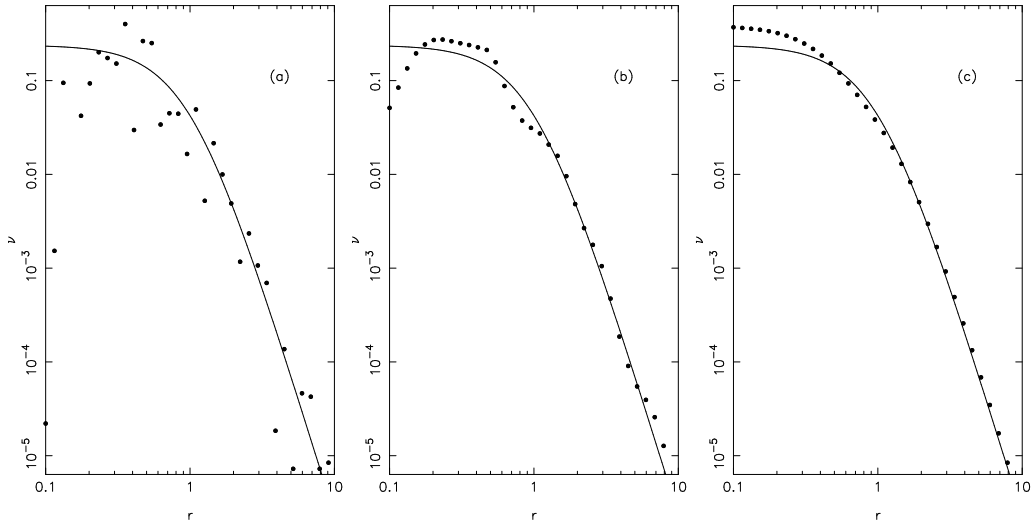

Fig. 2. Maximum penalized likelihood estimates of $\nu(r)$ for a sample of 300 positions drawn from the Plummer density law in projection, and three different values of $\lambda$.

A number of problems in galactic dynamics can be expressed in the form of Abel integral equations. One example arises in the study of globular clusters, which are often modelled as spherical systems with isotropic velocity distributions, $f = f(E)$. Suppose we measure the surface density and line-of-sight velocity dispersion profiles of some set of stars in a globular cluster. Their intrinsic number density and velocity dispersion profiles follow from regularized algorithms like those given above; the projection integral for isotropic velocity dispersions is

$$\nu_p(r_p)\sigma_p^2(r_p) = \int_{r_p^2}^{\infty} \frac{\nu(r)\sigma^2(r)dr^2}{\sqrt{r^2 - r_p^2}}. \tag{17}$$

The gravitational potential is then given by the spherical Jeans equation

$$\frac{d\Phi}{dr} = -\frac{1}{\nu}\frac{d(\nu\sigma^2)}{dr} - 2\frac{\beta\sigma^2}{r} \tag{18}$$

with $\beta = 1$. The isotropic distribution function is the solution to

$$\nu(r) = 4\pi \int_{\Phi(r)}^{0} \sqrt{2\left[E - \Phi(r)\right]} f(E)dE, \tag{19}$$

again solvable via a regularized routine. Thus it is possible to stably recover $f(E)$ and $\Phi(r)$ for a globular cluster, without assuming that $f$ has a particular form, or that the potential is generated self-consistenly from the observed population of stars. Nonparametric confidence intervals can be derived for any quantity of interest by asking, for instance, the range of central densities of models whose projected properties fit the data with a given accuracy, and whose smoothness lies within certain bounds.

(The fact that isotropic models can always be constructed to fit line-of-sight velocity dispersion data also demonstrates the fallacy of drawing conclusions about the anisotropy of a globular cluster from such data. Such conclusions are only valid in the context of a particular model for the potential, as discussed in the next section. If one does not force mass to follow light, as in a single-component Michie-King model, one finds that small adjustments in $\Phi$ can "compensate" for features in the observed profiles that would otherwise be attributed to anisotropy.)

If the velocity data are discrete, and too sparse to permit the accurate construction of a velocity dispersion profile, an alternative technique is to compute a trial function $f(E)$ for some observed $\nu(r)$ and assumed $\Phi(r)$, using equation (19), then calculate the likelihood that the observed kinematical sample was drawn from this $f(E)$. This technique was used by Merritt & Tremblay (1992) to map the potential surrounding the giant elliptical galaxy M87, using the $\sim 45$ globular clusters with measured radial velocities. A very similar technique was used by Kulessa & Lynden-Bell (1992) to measure the mass of the Galaxy from a sample of $\sim 50$ satellites. In both cases, of course, the inferred mass model is valid only to the extent that the isotropy assumption is correct.

The Abel equation appears in many contexts outside of astronomy, and a wide variety of techniques have been developed since the mid 1960's for dealing with its inherent instability (e.g. Gorenflo & Kovetz 1966; Cremers & Birkebak 1966; Minerbo & Levy 1969; Anderssen 1976; Bendinelli 1991). It is distressing that astronomers have paid so little attention to this work. The explanation may be pedagogical : generations of astronomers have been urged, in King's (1981) words, to "always bring the theoretical quantities into the observational domain, rather than vice versa. You can convert the theoretical quantities with as high an accuracy as you like, whereas observational data are nearly always degraded by a conversion." Although this philosophy is certainly proper when the aim is to verify theoretical predictions, it is less appropriate as a blueprint for statistical inference, where the goal should be to "let the data speak for themselves."

## 3. RECOVERING ANISOTROPIC DISTRIBUTION FUNCTIONS

A much-studied inverse problem in galactic dynamics is the recovery of the *anisotropic* distribution function $f(E, L^2)$ of a spherical galaxy given measurements of the surface density, $\nu_p(r_p)$, and line-of-sight velocity dispersion, $\sigma_p(r_p)$, of some equilibrium population. Unlike the problem just discussed, the potential $\Phi(r)$ is here assumed known; it might be calculated, for instance, from the virial theorem, under the assumption that mass follows light, or it might be the assumed potential of some dark component. By assigning a functional form to $\Phi(r)$, the relation between the model $(f)$ and the data $(\nu_p, \sigma_p)$ becomes linear, which is the primary motivation for stating the problem this way. The cost is that very little can be inferred about the form of the potential aside from its normalization.

The mathematical relations between $f$, $\nu_p$, $\sigma_p$ and $\Phi$ are :

$$\nu_p(r_p) = 4\pi \int_{r_p}^{\infty} \frac{dr}{r\sqrt{r^2 - r_p^2}}$$
$$\int_{\Phi(r)}^{0} dE \int_{0}^{2r^2[E-\Phi(r)]} f(E, L^2) \left\{ 2[E - \Phi(r)] - L^2/r^2 \right\}^{-1/2} dL^2 \tag{20}$$

and

$$\nu_p(r_p)\sigma_p^2(r_p) = 4\pi \int_{r_p}^{\infty} \frac{dr}{r\sqrt{r^2 - r_p^2}}$$
$$\int_{\Phi(r)}^{0} dE \int_0^{2r^2[E-\Phi(r)]} f(E, L^2)\left\{2[E - \Phi(r)] - L^2/r^2\right\}^{1/2} dL^2. \quad (21)$$

The first integrals on the right hand side of equations (20) and (21) are projections along the line of sight, while the second and third are integrals over velocity space. Because of the triple integration, we might expect these equations to be strongly ill-posed with respect to the recovery of $f$ from $\nu_p$ and $\sigma_p$. On top of this, the problem as stated is *underdetermined*: there are generally many $f$'s consistent with a given set of observed profiles and a given $\Phi$. The nature of the indeterminacy was elucidated by Binney & Mamon (1982), who showed that the projected velocity dispersion profile contains only enough information to determine the intrinsic second moments of $f$, i.e., $\sigma_r(r)$ and $\sigma_t(r)$, the radial and tangential velocity dispersions. Explicit examples of this degeneracy have been presented by Dejonghe (1987a).

Three widely-used algorithms have been developed for recovering $f(E, L^2)$ from $\nu_p(r_p)$ and $\sigma_p(r_p)$ (Richstone & Tremaine 1984, 1988; Newton & Binney 1984; Dejonghe 1989a,b). These algorithms are usually described in terms of the schemes they adopt for discretizing and inverting the integral equations (20) and (21), e.g. "linear programming" vs. "quadratic programming," etc. Following the discussion above, we might guess that a more fundamental distinction between these algorithms is the way they deal with the ill-conditioned nature of the inverse problem. For instance, Richstone & Tremaine (1988) maximize the "collisionless entropy"

$$S = -\int C(f)d^3x\,d^3v \quad (22)$$

of their distribution function, where $C(f)$ satisfies $d^2C/df^2 \geq 0$, subject to the constraint that the projected model match the observed profiles on some grid of points $r_p$. In the terminology presented above, Richstone & Tremaine are using the "collisionless entropy" (22) as a regularization functional to stabilize the inversion. (Some unregularized solutions obtained with their algorithm are presented in Richstone & Tremaine 1985.) Because the problem is mathematically underdetermined the smooth solutions returned by their algorithm will in general depend strongly on their definition of $S$. Dejonghe (1989a,b) renders the inversion stable by representing $f$ via a truncated basis set,

$$f(E, L^2) = \sum_{\alpha,\beta} c_{\alpha,\beta}(-E)^{\alpha+\beta-3/2}L^{-2\beta}, \quad (23)$$

and choosing the $c_{\alpha,\beta}$ to minimize the mean square deviation of the projected $f$ from the data. In this way, he always finds a single, "best fit" model. As long as the number of basis functions is small (Dejonghe adopts, typically, $\sim 20$), the solution is guaranteed to be smooth. However Dejonghe's scheme does not deal in a very flexible way with the indeterminacy of $f$: among the many possible smooth solutions, his algorithm simply chooses the one that is most easily represented by his particular basis set. Newton & Binney (1984) use the Richardson-Lucy algorithm to solve for $f$, given a smooth initial guess $f_0$. [1] In effect, Newton and Binney embody their preconceptions about $f$ within the initial guess, and iterate toward a solution that is both more in accord with the observed profiles, and less smooth. But again, because the problem as posed is mathematically underdetermined, their final solution can depend strongly on their initial guess.

Note that these three algorithms will generally produce different solutions $f(E, L^2)$ even given the same input functions $\nu_p(r_p)$, $\sigma_p(r_p)$ and $\Phi(r)$. If the modeller's goal is simply to find *one* distribution function that is consistent with the data, this objection is not a serious one. However if the goal is to estimate the gravitational potential $\Phi(r)$, these algorithms break down entirely. At best, they can sometimes rule out certain $\Phi(r)$, by showing that no nonnegative $f$ consistent with the data exists in that potential (e.g. Dressler & Richstone 1990; Saglia et al. 1992); but they can never make useful statements about the relative probabilities of *different* potentials in which positive $f$'s exist. This shortcoming is quite severe in practice, since the range of potentials consistent with a specified surface brightness and velocity dispersion profile can be amazingly large (e.g. Merritt 1987).

Clearly it would be useful to estimate $f$ with an algorithm that begins from a statement of the problem that is mathematically determined (though probably still ill-posed in the Riemann-Lebesgue sense). One way to do this is to relate $f$ to the "projected distribution function,"

$$\nu_p(r_p, v_p) = \int dz \int \int dv_x dv_y f \left[ v_r^2 + v_t^2 + 2\Phi(r), r^2 v_t^2 \right], \qquad (24)$$

the joint distribution of projected positions and line-of-sight velocities; here $z$ is parallel to the line-of-sight, and $v_x$ and $v_z$ are the velocity components in the plane of the sky. Just as the lowest moments over $v_p$ of $\nu_p(r_p, v_p)$ uniquely determine – in a specified potential – the lowest velocity moments of $f$ (Binney & Mamon 1982; Merrifield & Kent 1990), so the complete projected distribution function uniquely determines $f(r, v_r, v_t)$ (Dejonghe & Merritt 1992). Of course, the amount of observational material required to nail down $\nu_p(r_p, v_p)$ is rather large : thousands of individual positions and velocities in the case of a discrete sample, or large numbers of accurate "line profiles" in the case of integrated spectra. However, by stating the problem in a mathematically determined way, one can *always* – regardless of sample size or quality of data – begin to make statements about the relative probabilities of different solutions, statements that are impossible to make using algorithms based on surface brightness and velocity dispersion profiles alone.

Equation (24) can be solved with a regularized, product-integration scheme similar to the one presented above for the solution of Abel's equation (Merritt 1992a). The major technical problem is translating a general $f$, specified on a grid in $(E, L)$ space, into a projected distribution function, specified on a grid in $(r_p, v_p)$. One simple way to accomplish this is to assume that $f$ is constant on small "patches" in $(E, L^2)$ space. The projection of any such patch is

---

simply

$$f_{patch} \times \int A(r, r_p, v_p)dz$$

where $A(r, r_p, v_p) = \int\int dv_x dv_y$ is the area at $r$ which the $(E, L^2)$ patch occupies in $(v_x, v_y)$ space. The projected distribution function corresponding to $f$ is just the superposition of the projected patches. The regularization functional must take into account the bi-dimensionality of $f$; a reasonable form is

$$P(f) = \int\int \left\{ \lambda_1(E, L^2) \left[ \frac{\partial^2 f}{\partial E^2} \right]^2 + \lambda_2(E, L^2) \left[ \frac{\partial^2 f}{\partial L^{2^2}} \right]^2 \right\} dE dL^2. \qquad (25)$$

Figures 3 and 4 illustrate the recovery of the Plummer model distribution function from complete, continuous data (via a quadratic programming scheme), and from limited, discrete data (by maximization of the penalized likelihood). Even quite modest samples – only 300, in the case of Figure 4 – permit the recovery of $f$ with reasonable precision.

It would make sense to begin applying algorithms like this one to the analysis of globular clusters. One might begin by estimating $f$ and $\Phi$ through fitting of surface density and velocity dispersion profiles to a Michie-King model. Given this $\Phi$, one could then ask, using the algorithm just described, for the anisotropic distribution function $f(E, L^2)$ implied by the full, joint distribution of stellar positions and velocities. If the Michie-King family is a good description of the globular cluster, these two $f$'s should not be significantly different.

## 4. ENTROPY

Two views of entropy are current in the astronomical literature. One view defines entropy as a measure of the degree of "evolution" of a collisionless stellar system away from some initial, well-defined state (e.g. Shu 1978; Tremaine, Hénon & Lynden-Bell 1986; White 1987). According to a theorem of Tolman (1938), if the distribution function of a collisionless system is known precisely at some time $t_1$, its value at some later time $t_2$ satisfies

$$S(t_2) \geq S(t_1) \qquad (26)$$

where $S$ is any functional of the form

$$S(\overline{f}) = -\int C(\overline{f})d\mathbf{x}d\mathbf{v}; \qquad (27)$$

$\overline{f}$ is the "coarse-grained" phase space density, and $C(x)$ is any "convex" function of $x$ (i.e. any function such that $d^2C/dx^2 \geq 0$). Tremaine, Hénon & Lynden-Bell (1986) call $S$ an "H function," and argue that it plays a role similar to that of Boltzmann's entropy in a collisional gas. On the basis of this argument, Richstone & Tremaine (1988) adopted the collisionless entropy as an appropriate regularization functional in their algorithm for finding $f(E, L^2)$. However this interpretation of $S$ has been criticized by Dejonghe (1987b), Sridhar (1987), Binney (1987) and others on the basis that Tolman's theorem cannot be used to establish that $S$ increases monotonically with time, like Boltzmann's entropy, but only that an initially, completely specified $f$ will phase-mix into an $f$ with finer and finer structure.

The second view of entropy is most strongly associated with E. T. Jaynes (1957), who was concerned, as we are here, with the problem of statistical estimation :

> ...in making inferences on the basis of partial information we must use that probability distribution which has maximum entropy subject to whatever is known. This is the only unbiased assignment we can make; to use any other would amount to arbitrary assumption of information which by hypothesis we do not have.

Jaynes derived – on the basis of probabilistic, not physical, arguments – an expression for the "entropy" that should be maximized, subject to any observational constraints, when estimating the form of an unknown frequency function $f$. In Dejonghe's (1987b) formulation, if $f_p$ is an *a priori* estimate of $f$, the Jaynes entropy may be written

$$S(f) = -\int f \left[\ln\left(\frac{f}{f_p}\right) - 1\right] d\mathbf{x}\, d\mathbf{v}. \tag{28}$$

In the absence of prior knowledge about the relative likelihoods of different $f$'s, $f_p$ is just a constant, and the Jaynes entropy is identical to the Boltzmann entropy. Following Jaynes (1957) and Dejonghe (1987b), when estimating $f$ for a galaxy from data that contain too little information to determine $f$ uniquely, we should find that model which is both consistent with the data, and maximizes the "entropy" (28).

It was shown above that – in a spherical stellar system – there are ways of casting the determination of $f$ as a problem that is, at least mathematically, fully determined by the data. Thus we might be tempted to think that entropy is left with no role to play in the modelling of such systems. Of course this would be too narrow a view. By interpreting our observations in terms of a simple spherical model, we are assuming information that we may not have : that the observed stellar system is not elongated along the line of sight; that departures from equilibrium are small; that rotation is negligible; etc. The perfect match between the information content of $f(E, L^2)$ and $\nu_p(r_p, v_p)$, in an assumed $\Phi(r)$, vanishes once we admit the possibility of these complicating factors. Thus – following Jaynes – we ought to model galaxies by calculating the entropy of every possible dynamical state that appears, in projection, to be consistent with the data. This is a tall order, not likely to be realizable with a computer code of finite size. Even within the context of our simple spherical model, however, entropy may have a role to play in eliminating the milder indeterminacy resulting from noise or incompleteness of data. Jaynes (1984) has argued that regularization schemes for the solution of unstable inverse problems should always have a basis in entropy, whether the associated mathematical problem is determined or underdetermined. It would be fruitful to explore this suggestion systematically, and to test whether inversion or estimation schemes based on a "maximum entropy" smoothing functional are systematically better than the rather *ad hoc* ones described above. This question has been the subject of some spirited debates in the literature (e.g. Titterington 1984 vs. Skilling 1984).

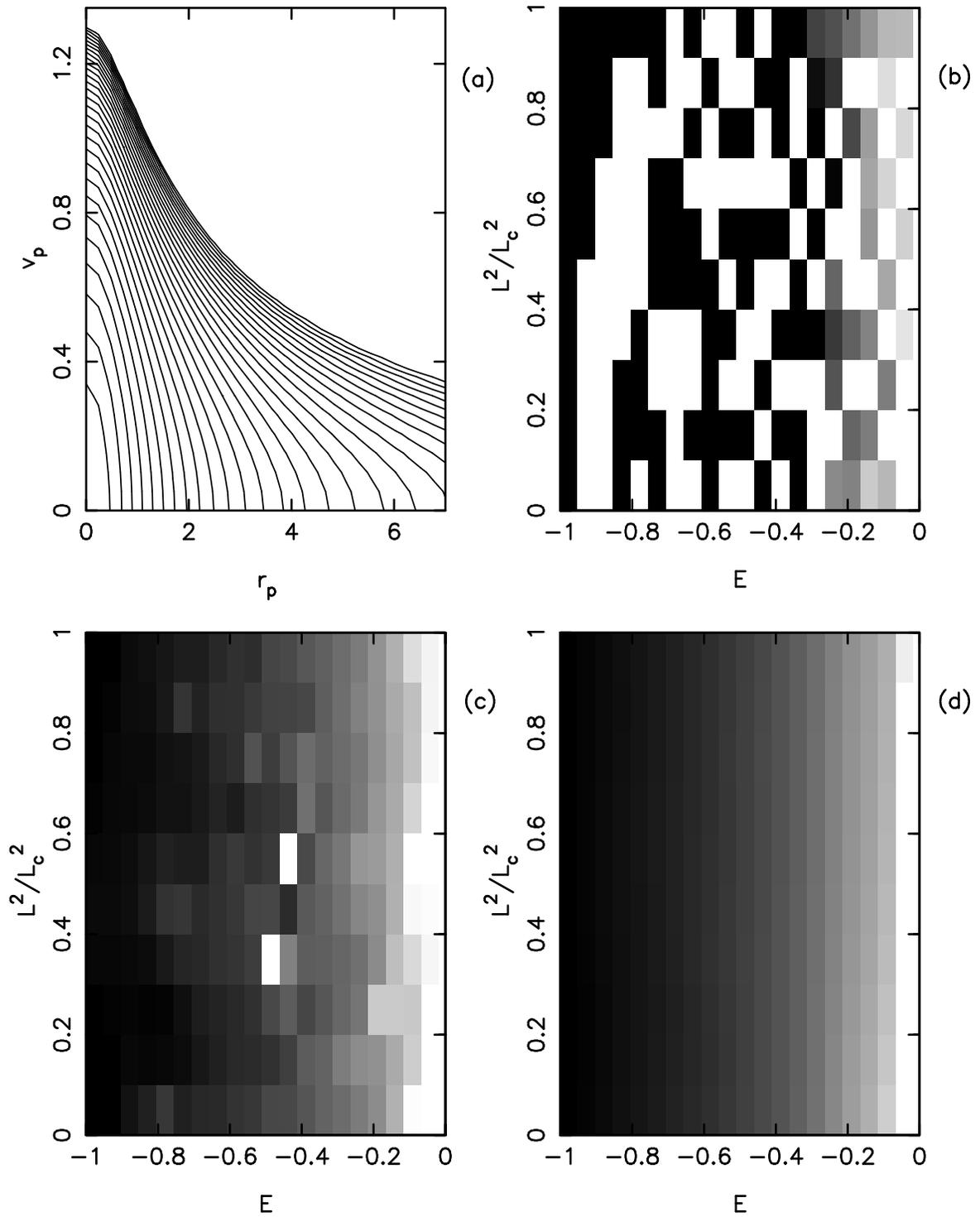

Fig. 3. Regularized inversion of the integral equation (24). (a) Projected distribution function of an isotropic Plummer model. (b) Phase space density resulting from direct inversion. (c) and (d) Regularized inversions, with two different values for $\lambda$. Negative values of $f$ are indicated by blanks.

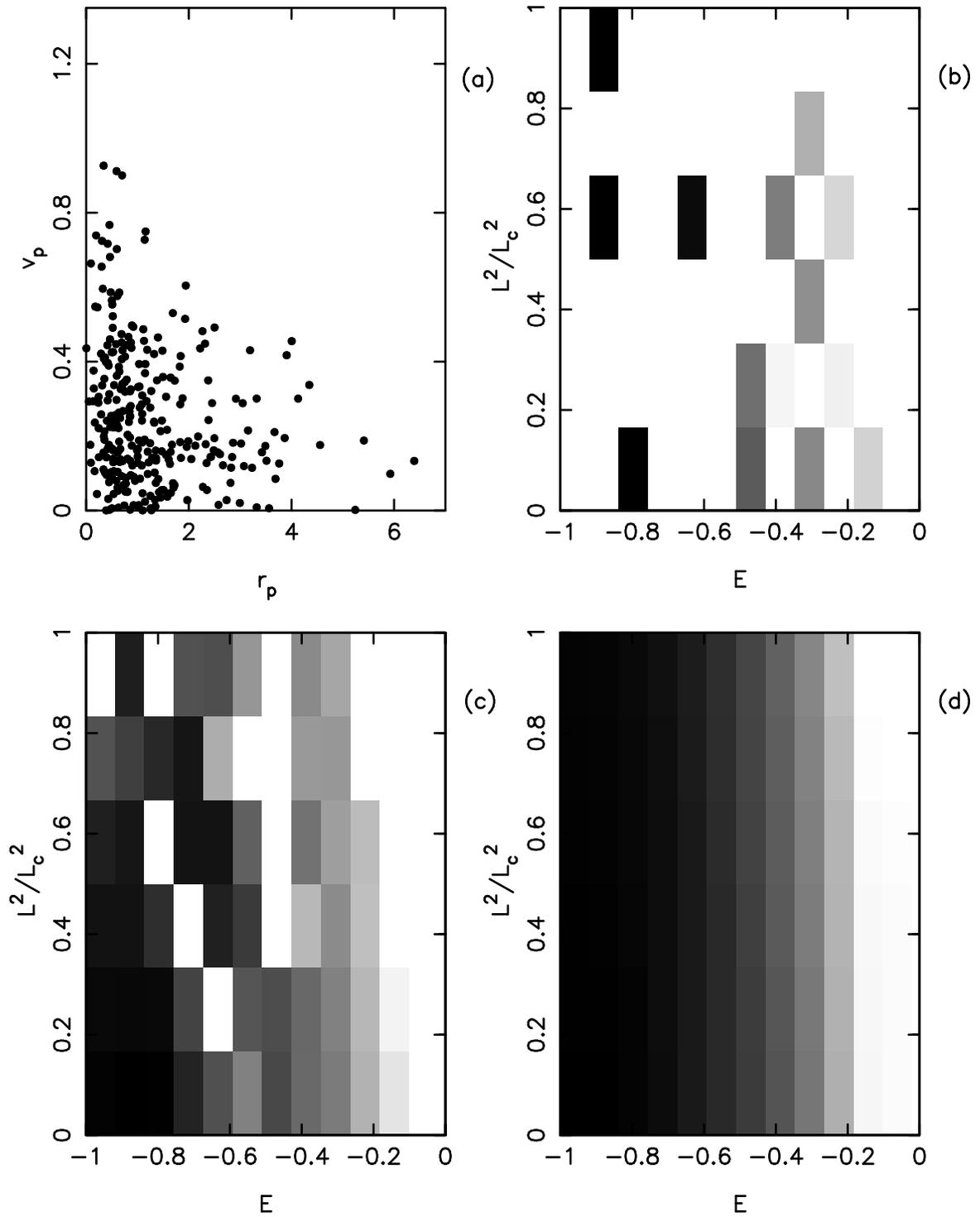

Fig. 4. Maximum likelihood estimation of $f$ from a discrete ($N = 300$) data set generated from the isotropic Plummer distribution function. (a) Data. (b) – (d) : Penalized optimization, with three different values for $\lambda$. The inferred $f$ in (d) is close to isotropic, like the true $f$.

# 5. POTENTIAL ESTIMATION

The problem discussed above, of inferring $f(E, L^2)$ from line-of-sight velocity data in an assumed potential $\Phi(r)$, is a natural one to attempt from a mathematical point of view, since $f$ is related to the data via a linear operator. This problem has much in common with the old "self consistency" problem of stellar dynamics, in which one attempts to find a stationary distribution of orbits that reproduces a specified density profile. However the analogy is misleading, since we rarely know the functional form of the potential in which the observed stars move. One could estimate the potential from the virial theorem,

$$\langle v^2 \rangle = \langle \mathbf{r} \cdot \nabla \Phi \rangle, \tag{29}$$

with $\langle v^2 \rangle$ equal to three times the line-of-sight mean square velocity. But because the virial theorem is an integral constraint, it tells us only about the normalization of $\Phi(r)$, and nothing about its functional form. Given the likely existence of dark halos, not to mention nuclear black holes, it would be foolish to assume that mass follows light anywhere in a galaxy. An estimate of the spatial dependence of the potential can obviously only be made if the data themselves contain some information about the variation of stellar velocities with position. Since we can not hope to follow individual stars along their orbits, the best we can do is to compare the velocities of *different* stars at the same time, and ask what the spatial gradient in the velocity implies about the gradient of the potential. This problem is relatively simple when the orbital motion is one dimensional, as in the Oort problem (e.g. Kuijken 1991). It becomes much harder in the spherical case, since different stars can be on orbits of different shapes. The amount of information required to uniquely constrain $\Phi(r)$ is therefore much larger.

The most general form for a distribution function describing a nonrotating spherical system is $f = f(r, v_r, v_t)$, where $v_r$ and $v_t$ are the radial and tangential velocities. Suppose that one could somehow measure the three quantities $r$, $v_r$ and $v_t$ for each of a large set of stars in a spherical galaxy. It would then be possible to compute the phase space density $f(r, v_r, v_t)$ directly, by counting stars in tiny phase-space volumes. One could then ask for the unique function $\Phi(r)$ such that

$$f(r, v_r, v_t) = f\left[v_r^2/2 + v_t^2/2 + \Phi(r), r^2 v_t^2\right] = f(E, L^2), \tag{30}$$

i.e. the potential for which $f$ was consistent with Jeans's theorem. For instance, one could construct a single curve in the three dimensional phase space $(r, v_r, v_t)$ along which $f$ and $L = r v_t$ were constant, and use the fact that $E = v_r^2/2 + v_t^2/2 + \Phi(r)$ must also be constant along this curve to read off $\Phi(r)$. Thus a direct determination of $f(r, v_r, v_t)$ is tantamount to a determination of $\Phi(r)$.

Unfortunately, the information contained within the joint distribution of projected positions and line-of-sight velocities, $\nu_p(r_p, v_p)$, is much less than the information required to specify the three-dimensional function $f(r, v_r, v_t)$. This does not imply that $\Phi(r)$ is inaccessible, however, since it is clear that complete knowledge of $f(r, v_r, v_t)$ *over*constrains the potential : there are many independent curves in $(r, v_r, v_t)$ space along which we could carry out the exercise described above, and all of these must give the same $\Phi(r)$. We might therefore expect that we can uniquely determine $\Phi(r)$ with less information than is contained within $f(r, v_r, v_t)$.

This turns out to be true: for instance, the three lowest moments of $f$, i.e. $\nu(r)$, $\sigma_r(r)$ and $\sigma_t(r)$, yield $\Phi(r)$ via the Jeans equation (18). The crucial – and to some extent, still unanswered – question is: what sorts of kinematical information, seen *in projection*, are required to uniquely constrain $\Phi(r)$? This is a difficult question to answer in general because of the highly nonlinear relation between the data and $\Phi$. The surface brightness and line-of-sight velocity dispersion profiles are grossly insufficient for this purpose (Dejonghe & Merritt 1992). Measuring two independent velocity dispersions in the plane of the sky, via proper motions, uniquely determines the potential, since in a spherical system the proper motion velocity dispersions can be deprojected to yield the intrinsic velocity dispersions $\sigma_r(r)$ and $\sigma_t(r)$ (Leonard & Merritt 1989). This technique is beginning to be applied to some nearby globular clusters (Rees & Cudworth 1992), and should eventually tell us whether the mass-to-light ratio in globular clusters is a strong function of radius. However most galaxies are too distant to permit the measurement of velocity components that do not lie along the line of sight.

Since the most kinematical information that one can hope to obtain for a distant galaxy is the joint distribution of projected positions and line-of-sight velocities, or what was called above the "projected distribution function," it would be nice to know exactly what this function implies about the dynamical state of a hot stellar system. In spite of a fair number of papers having been written on this subject (e.g. Merritt 1987; Merrifield & Kent 1990; Kent 1991; Dejonghe & Merritt 1992; Merritt 1992a), the answer is not really known. Kent (1991) showed that the projected distribution function imposes a formally infinite set of nonlinear integral constraints on $\Phi(r)$, the lowest order of which is the usual virial theorem. Dejonghe & Merritt (1992) developed a machinery for constructing the most extreme potentials consistent with a given set of observed velocity moment profiles, and showed that the requirement that the intrinsic moments of $f$ be nonnegative further constrains the form of the potential. the fabout the for scale-free Although this work is suggestive, a convincing proof that the projected distribution function uniquely constrains (or, more likely, overconstrains) the potential in a spherical system has not yet been presented. If this hypothesis is true, it implies that both the potential and the distribution function of a spherical system are in principle accessible using line-of-sight velocity data. This conclusion seems now to be generally accepted by observers and model builders, and it will likewise be assumed in what follows. A formal proof would be nice to have, however.

It should be emphasized that basing estimates of $\Phi$ on $\nu_p(r_p, v_p)$ does not necessarily require a huge increase in the amount or quality of kinematical data beyond what is already available (though more and better data are always an advantage). The important thing is to look at the data in the proper way. For instance, a finite sample of positions and velocities can always – whatever its size – be viewed as a discrete approximation to the projected distribution function, rather than as a set of numbers from which to compute the dispersion profile. Similarly, one can always ask for the smooth function $N(v_p)$ whose convolution with some stellar template best describes an integrated galaxy spectrum, rather than simply assuming a Gaussian. (The deconvolution of stellar absorption line spectra is an interesting inverse problem in its own right; see Bender 1990 and Rix & White 1992.) As long as an attempt is made to glean more information from the data than its low order moments, one can begin to make valid statements about the most likely $f$ and $\Phi$. The "best" way to get this additional information from the data is open to debate. Van der Marel & Franx (1992) advocate representing the line-of-sight velocity distribution in terms of a Gram-Charlier series,

$$N(v_p) = e^{-v_p^2/2\sigma^2} \sum_{j=0}^{n} a_j H_j(v_p/\sigma), \qquad (31)$$

with $H_j$ the Hermite polynomials. The primary justification is that line profiles from theoretical models are often well approximated by Gaussians. Gerhard (1992) then suggests using the coefficients of the Gram-Charlier expansion as diagnostics for the degree of velocity anisotropy, and hence the potential. While generally superior to an expansion in terms of moments over $v_p$, the Gram-Charlier series is inefficient at representing the tails of a distribution (Kendall & Stuart 1958), and is ill suited to discrete data (e.g. Shenton 1951). Other parametrized families of distributions based on the normal distribution have similar shortcomings (Tapia & Thompson 1978). In addition, it is hard to see how to make efficient use, in a scheme like Gerhard's (1992), of the fact that line profiles at different radii are coupled through the Boltzmann equation.

Here again, a completely nonparametric approach is probably best. In any assumed potential $\Phi(r)$, one can find the unique, smooth distribution function $f(E, L^2)$ from which the data were most likely to have been drawn, using an algorithm like the one described above. This exercise can be repeated for different assumed potentials, until the single potential that best fits the data is found. To the extent that the mathematical problem is a determined one, and the information content of the observations has not been too strongly compromised by the data reduction process, this approach will always yield a unique, "best fit" $f$ and $\Phi$. However the confidence intervals associated with each function, and therefore the ability to distinguish in a significant way between different solutions, will depend strongly on the amount and quality of the data. Work to date using flexible nonparametric algorithms (Merritt & Saha 1992; Merritt 1992a; Merritt & Tremblay 1992) suggests that the amount of kinematical information required to place interesting constraints on the radial form of the potential can be very large. In the case of integrated spectra, ruling out a constant M/L model for a galaxy with an isothermal halo requires accurate ($\lesssim 10\%$) measurements of the line profiles out to two or three effective radii (Merritt 1992a). Because of limitations imposed by the surface brightness of the night sky, such observations will be very difficult. In the case of discrete data, samples of several hundred or a thousand appear to be a practical minimum when discriminating between different exponents for the mass density falloff at large radii (Merritt & Tremblay 1992). Such a large number is not surprising given that a hundred or more discrete velocities are required to specify the line-of-sight velocity distribution at a single projected radius.

With both continuous and discrete data, the prospect appears somewhat brighter for placing constraints on the radial form of the potential near the center, where surface densities are high. For instance, Merritt & Saha (1992) find that the ~300 measured velocities in the Coma galaxy cluster are sufficient to place interesting upper limits on the core radius of the mass distribution. Much more work will be required before definitive estimates of the matter distribution will be available for a variety of real systems. However it is already clear that there is excellent justification for measuring many hundreds, even thousands, of radial velocities in individual systems, many more velocities than the hundred or so that suffice for the construction of the velocity dispersion profile. Such samples have already begun to appear for some globular clusters (e.g. Reijns et al. 1992) and for the Galactic bulge (te Lintel Hekkert & Dejonghe 1989).

The statements in the preceding paragraphs seem to contradict a large body of past work on potential estimation in hot stellar systems. In many of these studies, strong conclusions about the distribution of mass have been reached from kinematical samples much smaller than advocated above. For instance, a recent study of the mass of the Galaxy (Little & Tremaine 1987), based on a sample of just ten objects, concludes that an extended halo model can be effectively ruled out. A common practice in these studies is to assign low probabilities to dynamical models that seem *physically* implausible, whether or not those models are inconsistent with the data. Little & Tremaine (1987), for instance, chose to exclude models with strongly tangentially biased velocity distributions. In other studies, restrictions are placed on $f$ for reasons of mathematical or computational convenience only. When the underlying estimation problem is underdetermined, or the data are sparse, the resulting bias in the estimate of $\Phi$ can be enormous. Two examples from the recent literature provide illustrations. The & White (1986) attempted to find the most likely form of the mass distribution in the Coma galaxy cluster by fitting models to the projected velocity dispersion profile of the galaxies, obtained by binning galaxies in radius and taking second moments over $v_p$. Their preferred model had a mass distribution similar to that of the galaxies, with a roughly isotropic velocity distribution. Since a projected velocity dispersion profile contains too little information to uniquely determine $\Phi$ or $f$, The & White's recovery of a preferred mass model must have resulted from some feature in their algorithm that restricts the functional form of $f$. In fact the problem can be traced to their use of a simple parametrized form for the radial velocity dispersion profile $\sigma_r(r)$. A nonparametric study (Merritt 1987) confirms that an extremely wide range of mass models are *equally* consistent with the velocity dispersion data, as one would expect for an underdetermined problem.

A second example is the attempt by Kulessa & Lynden-Bell (1992) to infer the mass of the Milky Way galaxy using a sample of $\sim 50$ distant satellites with measured radial velocities and distances. By assuming an isotropic velocity distribution for the satellites, they found a most likely power-law mass model with a total mass of $1.3 \times 10^{12} M_\odot$ and power-law index of 2.4. They then tested the sensitivity of their result to the assumption of isotropy by postulating a family of constant-anisotropy distribution functions. They found that even mild tangential anisotropies ($\sigma_t/\sigma_r = 1.4$) implied an increase of about 50% in the most likely total mass. Kulessa & Lynden-Bell do not discuss the relative likelihoods of the isotropic and tangential models, but given the small size of their sample, they probably could not hope to distinguish between the two cases. Thus their conclusions about the mass of the Galaxy are strongly dependent on the assumed kinematics of their tracers.

The modest kinematical samples analyzed by Little & Tremaine (1987) and Kulessa & Lynden-Bell (1992) probably do not contain enough information to usefully constrain $f$ or $\Phi$ in the absence of fairly restrictive assumptions about the form of those functions. It is reasonable in these circumstances to assign low probability to models that seem physically implausible, whether or not the models can be shown to be incompatible with the data. But this means that their results, like the results of most work to date on potential estimation in hot stellar systems using kinematical tracers, are substantially model-dependent. Happily, this state of affairs should be short-lived, as much larger and more accurate kinematical samples begin to appear. In the near future, it should be possible to infer the distribution of matter around hot stellar systems with the same degree of accuracy that now applies to the mapping of dark matter around spiral galaxies.

## Acknowledgements

I thank C. Pryor and J. Sellwood for careful readings of the manuscript and comments which improved the presentation.